\documentclass{article}

\usepackage[english]{babel}

\usepackage[letterpaper,top=2cm,bottom=2cm,left=3cm,right=3cm,marginparwidth=1.75cm]{geometry}

\usepackage{amsmath}
\usepackage{graphicx}
\usepackage[colorlinks=true, allcolors=black]{hyperref}

\begin{document}
\begin{titlepage}
   \begin{center}
       \vspace*{1cm}

       {\LARGE \textbf{Systematic Mapping Protocol}}

       \vspace{2cm}
        {\Large Variability Management in Dynamic Software Product Lines for Self-Adaptive Systems}
            
       \vspace{1.5cm}

       \textbf{Oscar Aguayo, Samuel Sepúlveda}

       \vspace{5cm}
    
       Departamento de Ciencias de la Computación e Informática\\
       Universidad de La Frontera\\
       Temuco, Chile\\
       May 2022
            
   \end{center}
\end{titlepage}
\setlength{\parindent}{0cm}

\begin{abstract}
\setlength{\parindent}{0cm}
\noindent
\textbf{Context:} The Importance of Dynamic Variability Management in Dynamic Software Product Lines.

\textbf{Objective:} Define a protocol for conducting a systematic mapping study to summarize and synthesize evidence on dynamic variability management for Dynamic Software Product Lines in self-adaptive systems.

\textbf{Method:} Application the protocol to conduct a systematic mapping study
according the guidelines of K. Petersen.

\textbf{Results:} A validated protocol to conduct a systematic mapping study.

\textbf{Conclusions:} First findings show that it is necessary to visualize new ways to manage variability in dynamic software product lines.

\end{abstract}

\section{Introduction}

With the constant evolution of software due to technological changes and its diversification in their needs, there is an increasing number of self-adaptive software systems, whose main feature is the adaptation of the system to the different needs, either the customer or the environment \cite{Weyns2019}. Therefore, these approaches must provide system reconfiguration capabilities, giving the possibility to adapt at runtime to react to changes in the context \cite{Weyns2019}. In \cite{Hinchey2012}, states that facing these challenges has led to the development of several approaches to adapt to changing needs, including self-adaptive, agent-based, autonomous, emergent, and bio-inspired systems. Dynamic software product lines (DSPLs) can meet these needs by providing a conceptual framework for managing variability in these types of systems, generating runtime changes, and managing system variability according to the needs of the environment \cite{Quinton2021}.\\

This paper aims to present the definition of a protocol to systematically map the evolution of variability management in Dynamic Software Product Lines during the reconfiguration process of self-adaptive systems between the years 2010 and 2021. The report is structured as follows. Section 2 describes the research method conducted, including the definition of the search string, the search process, and the criteria for inclusion and exclusion of articles. Finally, Section 3 presents the conclusions and future work.

\section{Protocol definition}
 conducting a systematic mapping study (SMS) according to the guidelines defined by Petersen \cite{PETERSEN20151}.

\subsection{Need}
This work is the initial part of a proposal that seeks to build and develop an architecture to manage variability in DSPL during the software reconfiguration process for self-adaptive systems, where it is necessary to know in detail the proposals that exist today in the area, as this will allow:
\begin{itemize}
\item Identify the requirements to generate a software reconfiguration while maintaining a level of stability in the running system during the process.
\item Know which technologies, tools, approaches, or others are used during the system reconfiguration process in DSPL and understand the justifications for use in each case.
\item Avoid activities or processes already performed by other authors.
\item Identify existing challenges in the area of reconfigurations for DSPL.
\end{itemize}

The results of the systematic mapping are expected to serve the SPL and eventually DSPL research community as a synthesis of what has happened in the last 11 years in run-time reconfiguration architectures for self-adaptive systems, both at the theoretical-practical and bibliometric levels.
\subsection{Objective}
The aim of this work is to collect several proposals of architectures or design patterns to manage the reconfiguration process in DSPL for self-adaptive systems present in the literature. It will seek to analyze various proposals of constraints in the reconfiguration process, main errors during the system adaptation process or also some notion regarding how to maintain system stability and reliability during reconfigurations, ensuring the consistency of the software with respect to the models and the running system \cite{Quinton2021}. For this purpose, a systematic mapping of the literature for the most relevant articles of the last 11 years will be carried out, with the intention of collecting as much data as possible on the proposals to analyze the methodological approach used to reconfigure the solution domain of a software in an execution environment. The final part of the paper is a synthesis and publication of results. It is expected that the publication of the systematic mapping will contribute to identify several challenges in the management of variability for the DSPL scientific community, as well as contribute to the integral development of the DSPL through the generation of a line of research.



\subsection{Research questions}
The research will be focused on obtaining background on how variability is managed in DSPL, main challenges and constraints used. The main question of our research refers to \textit{What are the main difficulties or errors in the proposed architectures for managing variability during the reconfiguration of Dynamic Software Product Lines in self-adaptive systems?}. In order to solve this research question, it will be subdivided into more specific questions, which are described in Table 1. 
\begin{table}[h!]
\centering
\begin{tabular}{|l|p{3cm}|p{3cm}|p{3cm}|} \hline
ID & RQ & Classification & Objective \\\hline
1 & What approach was used to apply constraints during software reconfigurations in DSPL? & Software Agents, Constraint language, Machine Learning, Metamodels, Reasoning engines, Static adaptations, Restrictions on feature modeling, Others. & Collect information on the type of practices used to maintain system stability during DSPL reconfigurations. \\\hline
2 & What methodologies are currently used to manage DSPL variability during reconfigurations? & -Proprietary architecture, MAPE-K, Agent-oriented software engineering, Third-party software & Identify how the variability model communicates with the running system to visualize reconfiguration changes \\\hline
3 & What are the current challenges in the management of DSPL? & & Highlight the main challenges to properly manage the dynamic variability of DSPL. \\\hline
\end{tabular}
\caption{\label{tab:widgets}Research questions}
\end{table}

\subsection{Publication Questions}
A set of publication questions (PQ) is included to complement the information collected and to characterize the bibliographic, demographic and bibliographic space. This includes the type of place where published papers and the number of articles per year, details are shown in Table 2.
\begin{table}[h!]
\centering
\begin{tabular}{|l|p{3cm}|p{3cm}|p{3cm}|} \hline
ID & PQ & Classification & Objective \\\hline
1 & What year was the article published? & 2010-2021  & Highlight how DSPL research has evolved over the years. Years with more publications \\\hline
2 & Where was the article published? & Journal, Congress & Identify the journals and congresses most interested in the study of the DSPL, analyzing the most predominant congresses, journals and publishers  \\\hline

\end{tabular}
\caption{\label{tab:widgets}Publication questions}
\end{table}
\subsection{Data Sources}
According to \cite{Kitchenham07guidelinesfor}\cite{BRERETON2007571} we consider the data sources detailed in Table 3, that are recognized among the most relevant in the Software Engineering community.

\begin{table}[h!]
\centering
\begin{tabular}{|l|l|} \hline
Library & URL \\\hline
ACM Digital Library & \href{https://dl.acm.org}{dl.acm.org}\\\hline
IEEE Xplore & \href{https://ieeexplore.ieee.org/Xplore/home.jsp}{ieeexplore.ieee.org} \\\hline
Science Direct & \href{https://www.sciencedirect.com}{sciencedirect.com} \\\hline
Springer Link & \href{https://link.springer.com}{springer.com} \\\hline
Wiley Inter-Science & \href{https://onlinelibrary.wiley.com/search/advanced}{onlinelibrary.wiley.com} \\\hline
\end{tabular}
\caption{\label{tab:widgets}Data Sources}
\end{table}

\subsection{Search String}
The search string has been constructed according to the steps defined by \cite{Kitchenham07guidelinesfor}, through the context and the research question a set of keywords has been extracted, then for each keyword a set of synonyms has been proposed to widen the search range.
Using PICOC \cite{Wiley2006}, the search string for the systematic mapping is constructed.
\begin{itemize}
    \item Population: Dynamic Software Product Lines (DSPL)
    \item Intervention: Variability management approaches for DSPL.
    \item Comparison: The comparison does not apply to our study because the RQs did not consider the comparison of the assembled documents with a specific DSPL variability management proposal.
    \item Outcome: Software reconfigurations.
    \item Context: Dynamic Software Product Lines and Self-Adaptive Systems.
\end{itemize}
The list of keywords and synonyms is listed as follows:
\begin{itemize}
\item \textit{adaptation / reconfiguration}
\item \textit{dynamic software product lines / software product lines}
\item \textit{self-adaptive systems / adaptive systems / evolution}
\end{itemize}

The final query string is described as follows: \\
\textit{("self-adaptive systems" OR "self-adapting systems" OR "autoadaptive systems" OR "self-adaptive" OR "self-adapting" OR "adaptive systems" OR “Self-adaptation” OR “Adaptation”) \\
AND \\
("dynamic software product line" OR "dynamic software product lines" OR "dynamic product family" OR "dynamic product families" OR "dynamic product line" OR "dynamic product lines" OR “Software product line” OR “software product lines” OR “product family” OR “product families” OR “product line” OR “product lines”) \\
AND\\
("system reconfiguration" OR "reconfiguration" OR “reconfiguration rules” OR “configuration rules” OR “adaptation rules” OR “evolution”)} \\

It is important to mention that the search string has been adapted for some search engines, due to the limitations of each one \cite{Gusenbauer2020}. However, each adaptation does not add or remove any filters.  Table 4 shows the total results of the search string used in the selected data sources. Due to the restriction in the search string of the \textit{Science Direct} data source, where a maximum of eight key terms is allowed, the following search string is used:\\

\textit{(“self-adaptive systems” OR “self-adaptive” OR “auto-adaptive systems” OR “adaptation”) AND
(“dynamic software product line” OR “dynamic software product lines” OR “software product lines” OR “software product line”)}

\begin{table}[h!]
\centering
\begin{tabular}{|l|l|} \hline
Library & Result \\\hline
ACM Digital Library & 765\\\hline
IEEE Xplore & 74 \\\hline
Science Direct & 485 \\\hline
Springer Link & 5511 \\\hline
Wiley Inter-Science & 1011 \\\hline
Total Result & 7846 \\\hline
\end{tabular}
\caption{\label{tab:widgets}Search string, total results (5 May 2022)}
\end{table}

\subsection{Search and Selection Process}
The search process starts with an automatic search in the selected electronic data sources using the search string defined in 2.6. The goal of this is to obtain the first collection of articles to be distributed among the researchers of the team. Each researcher has to independently filter this initial collection according to the following inclusion/exclusion criteria, which allow deciding whether an article in the collection is relevant or not for this study by reviewing only the title, abstract and keywords of each article. 

\begin{figure}[h!]
\centering
\includegraphics[width=0.8\textwidth]{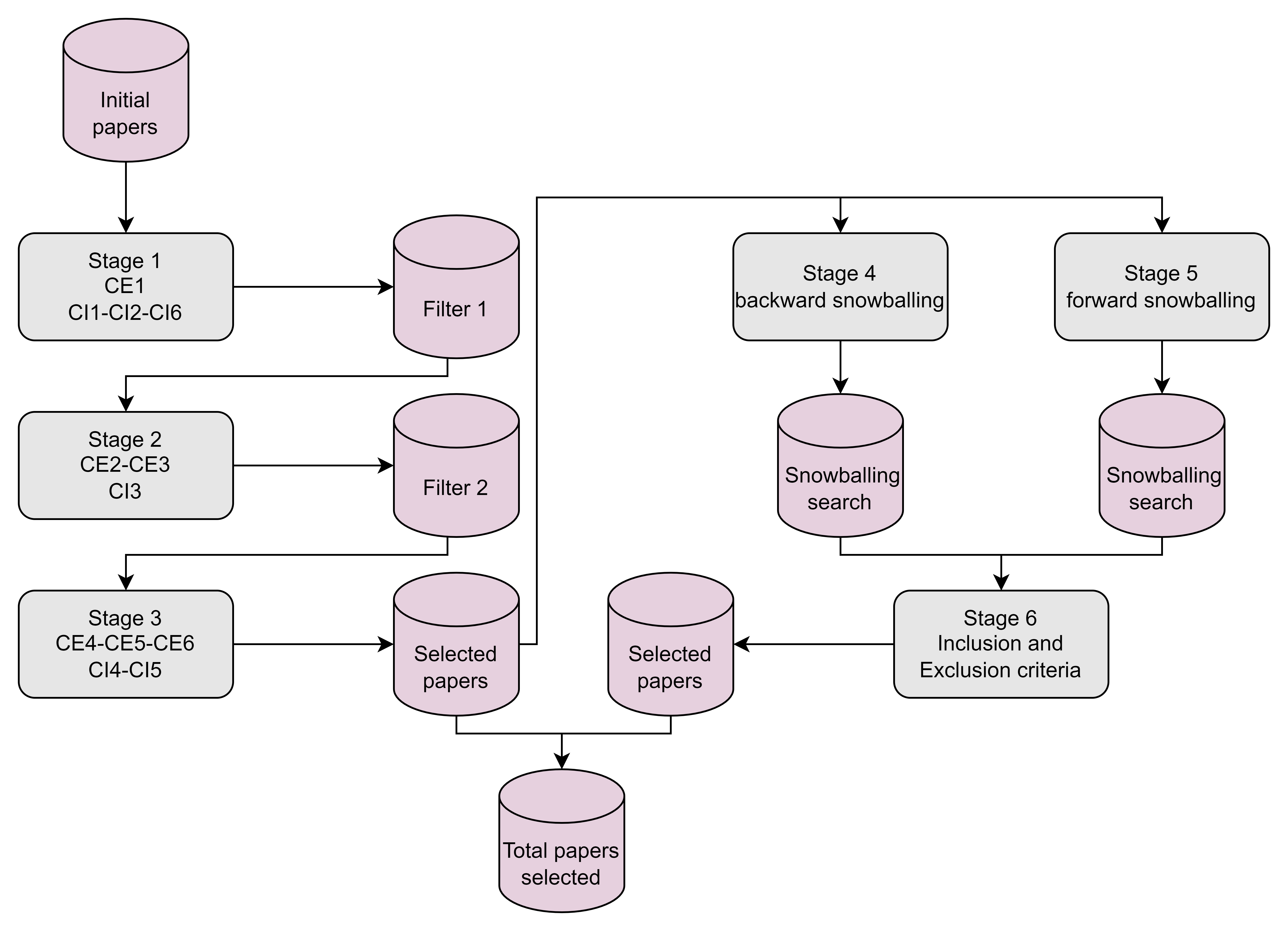}
\caption{\label{fig:searchPapers}Results from the search strategies}
\end{figure}

The first filter consists of applying the Inclusion (IC) and Exclusion (EC) criteria according to the following priority defined in Figure \ref{fig:searchPapers}. The definition of each IC and EC is shown in Tables 4 and 5.  
To avoid the exclusion of articles relevant to the research, we will occupy the snowballing search according to the guidelines proposed by Wohlin \cite{Wohlin2014}, in which the search range will be extended by reviewing the references of each article (backward snowballing) and the citations obtained by the articles (forward snowballing) already selected by the inclusion and exclusion criteria.

\begin{table}[h!]
\centering
\begin{tabular}{|l|p{10cm}|} \hline
ID & Criteria \\\hline
\textbf{IC1} & Articles published between 2010-2021.\\\hline
\textbf{IC2} & Papers written in English. \\\hline
\textbf{IC3} & {Type of paper:}\\ 
&
\begin{minipage} [t] {0.4\textwidth} 
      \begin{itemize}
      \item Proceeding.
      \item Journal.
      \item Conference Paper.
      \item Chapter LNCS (Lecture Notes in Computer Science).
     \end{itemize} 
    \end{minipage} \\ \hline
\textbf{CI4}    & Papers with more than one version, only the latest version will be included. \\\hline
\textbf{CI5} & Papers whose abstracts deal with Dynamic Software Product Lines or reconfigurations in Software Product Lines for self-adaptive systems.\\ \hline
\textbf{CI6} & {Topic:}\\ 
&
\begin{minipage} [t] {0.4\textwidth} 
      \begin{itemize}
      \item Computer Science.
     \end{itemize} 
    \end{minipage} \\ \hline
\end{tabular}
\caption{\label{tab:widgets}Inclusion criteria}
\end{table}

\begin{table}[h!]
\centering
\begin{tabular}{|l|p{10cm}|} \hline
ID & Criteria \\\hline
\textbf{CE1} & Articles written before 2010.\\\hline
\textbf{CE2} & Articles not related to Software Product Lines. \\\hline
\textbf{CE3} & Secondary researches. (If they exist, they will be added as related work).  \\ \hline
\textbf{CE4}    & Papers without access. \\\hline
\textbf{CE5} & Duplicate papers will be excluded.\\ \hline
\textbf{CE6} & {The following types of items will be excluded:}\\ 
&
\begin{minipage} [t] {0.4\textwidth} 
      \begin{itemize}
      \item Tutorials.
      \item Short Paper.
      \item Abstract.
      \item Poster.
      \item Paper in progress (incomplete).
      \item Book.
     \end{itemize} 
    \end{minipage} \\ \hline
\end{tabular}
\caption{\label{tab:widgets}Exclusion criteria}
\end{table}

\subsection{Resolving differences and avoiding bias}
In order to minimize possible differences in the conduct of the study, our protocol states the following decisions:
\begin{itemize}
      \item External validation of the search string and papers filtering.
      \item Publication of the protocol for public review on the arXiv platform.
      \item Collaborative definition of the SMS protocol and RQs.
      \item The researchers individually decide on the inclusion or exclusion of the assigned articles, chosen at random from those retrieved through a pilot selection.
      \item As a means of validation in the filtering and subsequent classification of the study, a concordance test based on Fleiss' Kappa statistic \cite{gwet2002kappa} will be performed. If $Kappa \geq 0.75$ then the criterion is sufficiently clear \cite{fleiss2013statistical}, otherwise, the criterion should be revised to obtain by its interpretation and application.
     \end{itemize} 
\subsection{Results and Reporting}
After having filtered the articles by inclusion and exclusion criteria, the relevant data will be extracted from the selected papers to answer the RQs and PQs.

The Meta-data collected for each article: (i) title, (ii) authors (each author), (iii) year of publication, (iv) publication type and classification, (v) approach used to generate reconfigurational constraints in DSPL, (vi) methodologies used to manage variability, (vii) challenges in DSPL management, (viii) results and future work.

\begin{figure}[h!]
\centering
\includegraphics[width=1\textwidth]{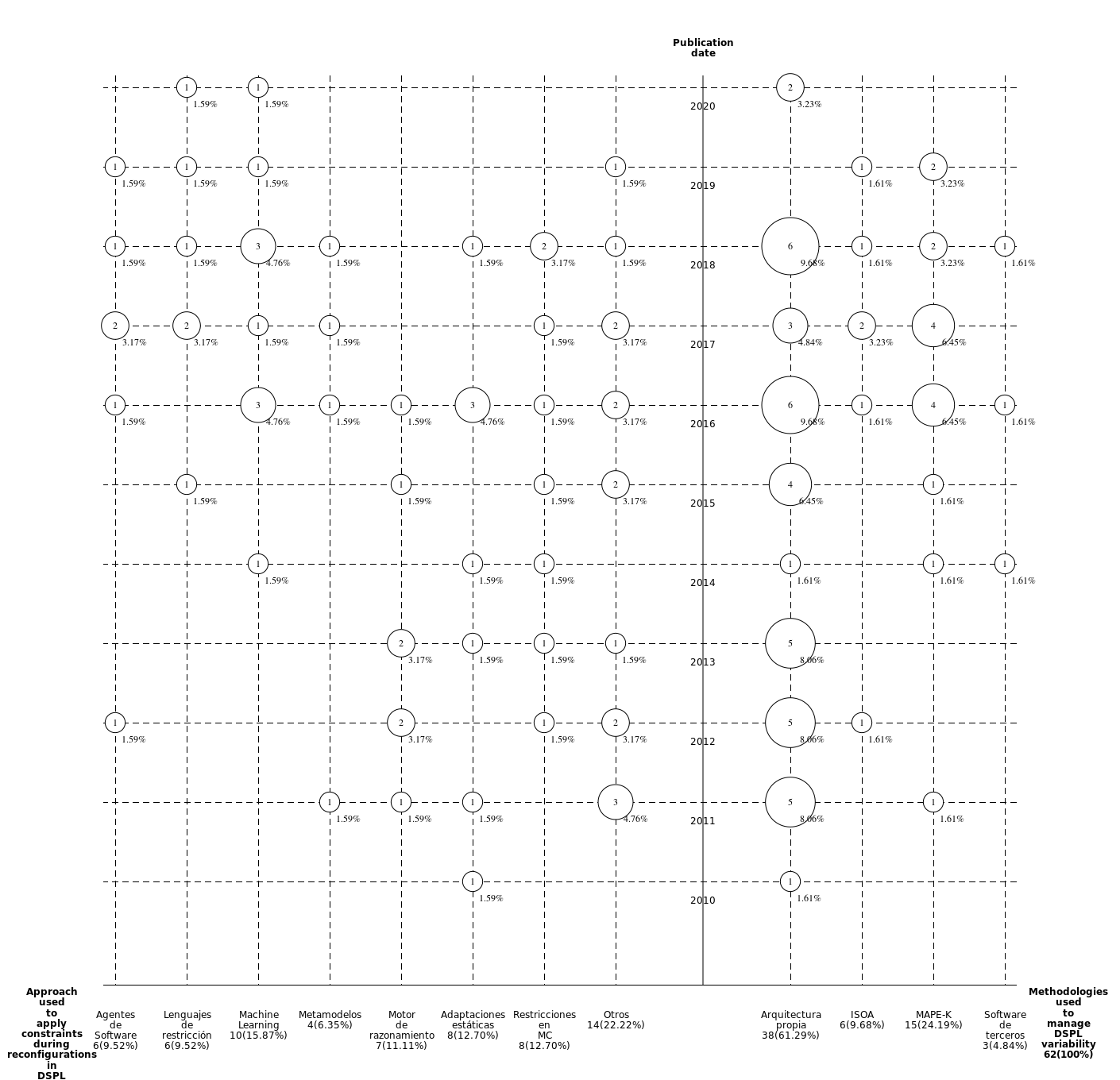}
\caption{\label{fig:reporte}Bubble Diagram Example}
\end{figure}

Another output will be a bubble diagram that graphically represents the number of documents found divided into categories. An example is shown in Figure \ref{fig:reporte}.

Presenting the results of this SMS considers 3 stages:
\begin{itemize}
    \item \textbf{Stage 1:} Publish the SMS protocol at Arxiv web platform.
    \item \textbf{Stage 2:} Publish the results of the SMS in an academic journal or at a conference focused on the topic of software engineering and DSPL.
    \item \textbf{Stage 3:} Present a proposal for dynamic variability management based on the problems encountered in the SMS.
\end{itemize}

The report with the results of phase 2 will be prepared with the structure recommended by \cite{Kitchenham07guidelinesfor}. The main sections are:
\begin{itemize}
    \item Introduction
    \begin{itemize}
        \item Context
        \item Aim and need
    \end{itemize}
    \item Background
    \begin{itemize}
        \item DSPL
        \item Dynamic variability
        \item Self-adaptive systems
    \end{itemize}
    \item Related work
    \item Methodology
    \item Results and discussion
    \begin{itemize}
        \item Answers to RQs and PQs
        \item Threats to validity
        \item DSPL Challenges
    \end{itemize}
    \item Conclusion
    \begin{itemize}
        \item Conclusions
        \item Future work
    \end{itemize}
    
\end{itemize}

\section{Conclusion and Future Work}
We presented a protocol definition of an SMS to summarize and synthesize the evidence over the last few years on variability management in DSPL for self-adaptive systems. The initial results show that a detailed review needs to be done on the different challenges pertaining to the variability management of self-adaptive systems using DSPLs and their impact on an eventual architecture proposal to manage such variability.

As future work we plan to define an architecture or methodology that allows managing variability dynamically for DSPL ensuring minimum quality criteria, such as usability and performance in each reconfiguration.

\bibliographystyle{IEEEtran}
\bibliography{IEEEabrv,referencias}

\vspace{12pt}

\end{document}